\documentclass[journal]{IEEEtran}

\usepackage[adjust,compress]{cite}
 
\usepackage{graphicx,amsmath,amssymb}
\usepackage{url}
\usepackage{subfigure}
\usepackage{color}

\DeclareMathOperator{\tr}{tr}
\DeclareMathOperator{\rank}{rank}

\newcommand{\LL}{\mathcal{L}}
\newcommand{\set}[1]{\mathrm{#1}}

\newcommand{\ie}{{\it i.e.\/}, }
\newcommand{\eg}{{\it e.g.},}

\providecommand*{\mrm}[1]{\mathrm{#1}}
\newcommand{\dotp}[2]{\langle #1,#2\rangle}

\newcommand{\lexp}[1]{\mathrm{e}^{#1}}

\newcommand{\Qm}{Q_{\mrm{min}}}
\newcommand{\Qz}{Q_{0}}

\newcommand{\Dp}{D_{\mrm{p}}}
\newcommand{\Dr}{D_{\mrm{t}}}

\newcommand{\We}{W_{\mrm{e}}}
\newcommand{\Wem}{W_{\mrm{em}}}
\newcommand{\Wm}{W_{\mrm{m}}}
\newcommand{\wa}{\mathbf{w}_{\mrm{\alpha}}}
\newcommand{\we}{\mathbf{w}_{\mrm{e}}}
\newcommand{\wm}{\mathbf{w}_{\mrm{m}}}
\newcommand{\rp}{\mathbf{p}_{\mrm{r}}}
\newcommand{\um}{\mathbf{u}}
\newcommand{\xm}{\mathbf{x}}

\newcommand{\ju}{\mathrm{j}}
\newcommand{\diff}{\mathop{\mathrm{\mathstrut{d}}}\!}

\newcommand{\vpsi}{\boldsymbol{\psi}}

\newcommand{\rP}{P_\mrm{rad}}

\newcommand{\RR}{\mathbb{R}}
\newcommand{\CC}{\mathbb{C}}
\newcommand{\HH}{\mathbb{H}}
\newcommand{\minimize}{\mathop{\text{minimize\ }}}
\newcommand{\maximize}{\mathop{\text{maximize\ }}}
\newcommand{\subjectto}{\mathop{\text{subject\ to\ }}}
\newcommand{\Ss}{\set{S}}

\newcommand{\vJ}{\boldsymbol{J}}

\newcommand{\vF}{\boldsymbol{F}}

\newcommand{\vr}{\boldsymbol{r}}

\newcommand{\he}{\hat{\boldsymbol{e}}}
\newcommand{\hh}{\hat{\boldsymbol{h}}}

\newcommand{\hy}{\hat{\boldsymbol{y}}}

\newcommand{\hr}{\hat{\boldsymbol{r}}}

\newcommand{\hphi}{\hat{\boldsymbol{\varphi}}}

\newcommand{\vE}{\boldsymbol{E}}

\begin{document}
%
%
%
%
\title{On methods to determine bounds on the Q-factor for a given directivity}

\author{B.L.G. Jonsson, Shuai Shi, Lei Wang, Fabien Ferrero and Leonardo Lizzi
\thanks{B.L.G. Jonsson, Shuai Shi and Lei Wang are with KTH Royal Institute of Technology, at the School of Electrical Engineering, Stockholm, Sweden, e-mail: ljonsson@kth.se}%
\thanks{F. Ferrero and L. Lizzi are with the Universit\'e C\^ote d'Azur, CNRS, LEAT, Sophia Antipolis, France}
\thanks{Manuscript received May 18, 2017.}}

\markboth{Jonsson+etal}%
{Jonsson \MakeLowercase{\textit{et al.}}: On methods to determine bounds on the Q_factor} 
%

\maketitle

\begin{abstract}
This paper revisit and extend the interesting case of bounds on the Q-factor for a given directivity {for a small antenna of arbitrary shape}.  A higher directivity in a small antenna is closely connected with a narrow impedance bandwidth.  The relation between bandwidth and a desired directivity is still not fully understood, not even for small antennas.  Initial investigations in this direction has related the radius of a circumscribing sphere to the directivity, and bounds on the Q-factor has also been derived for a partial directivity in a given direction.  In this paper we derive lower bounds on the Q-factor for a total desired directivity {for an arbitrarily shaped antenna} in a given direction as a convex problem using semi-definite relaxation techniques (SDR). We also show that the relaxed solution is also a solution of the original problem of determining the lower Q-factor bound for a total desired directivity. 

SDR can also be used to relax a class of other interesting non-convex constraints in antenna optimization such as tuning, losses, front-to-back ratio. 
We  compare two different new methods to determine the lowest Q-factor for arbitrary shaped antennas for a given total directivity. 
We also compare our results with full EM-simulations of a  parasitic element antenna with high directivity.
\end{abstract}

\begin{IEEEkeywords}
Antenna theory, antenna Q, directional antennas, antenna radiation pattern, miniature antenna, fundamental limitations.
\end{IEEEkeywords}

\section{Introduction}
{Small antennas with higher than normal directivity are particularly interesting for internet-of-things (IoT) applications where the expected bandwidth of the device is very narrow. As an example, in Europe, the entire ISM-band at 0.9 GHz has a relative bandwidth of 2.8\% corresponding to a half-power Q-factor at 70. The ISM-band at 0.868 GHz has an even smaller bandwidth. Moreover, most of the applications are only using a 1 MHz bandwidth channel since the amount of data to be transmitted is very limited \eg\ for a physical parameter monitored by a sensor or the location coordinates calculated by a GPS receiver. Thus, even higher Q-factor might be feasible for IoT applications from the bandwidth perspective. High gain of small devices may enable them to communicate at a lower power-level, which is essential to reduce the power consumption in their communication mode, and hence save battery or harvested energy. High directivity can also be utilized to reduce interference with neighboring devices, which is essential to reduce packet collision and avoid the retransmission process~\cite{Le+etal2016}, thus saving on the power resources at both the device and network level. Moreover, for wearable applications, directive antennas are more efficient since they reduce the energy absorbed by the body. Against this background it is interesting to revisit and extend the bounds on the bandwidth with respect to a given directivity. Utilizing stored energies we will show that it is possible to obtain Q-factor bounds for a given total directivity.}

Stored energy based physical bounds for arbitrarily shaped small antennas provide interesting and useful information about antenna design possibilities~\cite{Gustafsson+Nordebo2013,Gustafsson+etal2016a,Capek+etal2016b}. They set the outer boundaries of what is possible. Such information is helpful before starting the non-linear and complex process of designing antennas and as a benchmark on the antenna performance. The stored energy of an antenna determines the Q-factor which is closely related to the fractional impedance bandwidth performance of the antenna~\cite{Yaghjian+Best2005,Gustafsson+Jonsson2014}. To determine practical Q-factor bounds it is essential to also include the key desired constraints on the antenna design parameters. Such constraints can be on the far-field, tuning, front-to-back ratio, gain, directivity requirements and tolerances on design and/or feeding amplitude and phase. 

However, only a small fraction of these and other constraints results in a convex optimization problem~\cite{Gustafsson+Nordebo2013}. In this paper we show that the we can formulate lower bounds on the Q-factor for a total directivity in a given direction as a relaxed convex optimization problem. Indeed a class of the non-convex constraint to Q-factor bounds can be relaxed into a convex problem. 

The relation between the antenna size and its maximal directivity has frequently been investigated, and it is known that any directivity can be obtained~\cite{Oseen1922,Uzkov1946,Riblet1948,Hansen1981}. Harrington~\cite{Harrington1958} used $N$ spherical-modes to derive a limit for {\it normal} directivity as a function of size and Geyi~\cite{Geyi2003} used Q-factors for spheres to obtain another limit for the normal directivity as function of the radius of an enclosing sphere. Beyond those spherical-shape based size-limitations, it is also well known that one can obtain bounds on the Q-factor for an arbitrary shaped antenna, for a given partial directivity~\cite{Gustafsson+Nordebo2013}. However the extension of this latter bound to the total directivity results is a non-convex problem, and has thus been difficult to determine. 

It is therefore very interesting that the semi-definite relaxation methods~\cite{Lovasz1979,Goemans+Williamson1995} enables us to obtain lower bounds on the Q-factor for this class of problems. Indeed this technique also allows us also to introduce other interesting non-convex constraints, like tuning and losses.  A bound in~\cite{Gustafsson+Nordebo2013} determines the lowest possible Q-factor for a given \emph{partial} directivity in a given direction. This work is in the same direction: find the lowest possible Q-factor bound for a given \emph{total} directivity, in a desired direction. It rests on the observations~\cite{Oseen1922,Uzkov1946,Riblet1948,Hansen1981} that any directivity is theoretically possible given a sufficiently high $Q$-factor. High enough Q-factors make such antenna design solutions difficult to realize and to measure~\cite{Hansen2006,Yaghjian+etal2008}, it's thus more practically interesting to determine trade-offs between high directivity and as low as possible Q-factor. Earlier efforts in this direction has also been considered by \eg~\cite{Lo+Lee1966,Margetis+etal1998,Gustafsson+Nordebo2013} in different contexts. We illustrate this type of lower Q-bounds for two types of shapes, and compare the results with a full-wave simulated highly-directive antenna. An overview of small antennas with high directivity can be found in~\cite{Pigeon+etal2014}.

There are advantages if the investigated optimization problem can be formulated as a convex problem: Such problems have a unique minimum, furthermore there are fast numerical methods to find both the minimum and minimizer see \eg~\cite{Grant+Boyd2014}. These methods are so effective that sometimes the problem is considered `solved' as soon as it has a convex formulation. This is not quite that straightforward for antenna designs, but knowing bounds on the Q-factor certainly helps in defining antenna performance goals. 

If the investigated constraint to the Q-factor optimization makes the problem non-convex, it is in general much harder to solve the problem. 
Tools like tracing multiple Lagrange parameters over some sub-domain genetic algorithms and exhaustive searchers 
are in general considerably slower and the problem might not have a unique minimum. 
In the case of Q-factor bounds we find that almost all constraints on the antenna parameters are linear or quadratic in the current density. 
This class of optimization problem is called quadratically constrained quadratic program (QCQP).  It is known that this class includes NP-hard problems see \eg~\cite{Goemans+Williamson1995}, and that such problems are in general not convex. 

Successful convex methods to estimate a lower bound for the minimum of a QCQP problem are the semi-definite relaxation method (SDR)~\cite{Lovasz1979,Goemans+Williamson1995,Luo+etal2010} and the reformulation-linearization technique see \eg~\cite{Anstreicher2009}, see also~\cite{Sun+Dai2016}. SDR, which is utilized in this paper, has recently been used to solve several interesting problems, ranging from beamforming in communication networks, optimal power in power systems, to phase-reconstruction~\cite{Bengtsson+Ottersten2001,Luo+etal2010,Lavaei+Low2012,Candes+etal2015}. SDR converts a QCQP-problem to a semi-definite programming problem see \eg~\cite{Vandenberghe+Boyd1996}. The idea rests on a trace-operator identity that helps to reformulate the original problem into a linear relaxed problem. The original problem is thus formulated into an optimization of the trace of matrix products, usually also with constraints in terms of traces.

Trace-based methods provide a powerful tool to rewrite large classes of optimization problems. It has been shown that a class of data mining and machine learning problems can be formulated as trace-optimization problems in order to extract a set of small dimension out of a very large data sets see \eg~\cite{Kokiopoulou2011}. This latter type of trace-optimization problems tends to be quadratic in their unknown, as a difference to the SDR problems investigated here. But it is interesting to observe how the trace is utilized to obtain a low-dimensional solution in both cases.  

The numerical implementation of the relaxed problem gives a lower bound to the QCQP. The question naturally arises under what circumstances that the minimizer of the relaxed problem also solves the original numerical problem, for the distance between solutions to the SDR and to the original problem see~\cite{Huang+Zhang2007,Luo+etal2010}. 
One result shown in this paper is that the relaxed Q-factor bound for the total directivity in a given direction is tight. That is the solution to the relaxed problems for small antennas give the minimum to the original problem. This is also illustrated by comparison with a non-linear solver which is also described in the paper. 

The rest of this paper consists of five sections. Section II defines the Q-factor and certain key antenna parameters. Section III formulates the minimization problems for the Q-factor with different directivity constraints and the use of the SDR technique. Section IV introduces a non-linear eigenvalue based method to estimate the minimum Q-factor, and Section V illustrates a few numerical examples, in which lower bounds on the Q-factor for the different methods are investigated. Comparison with a simulated high-directivity antenna is also illustrated. Section VI concludes the paper. 

\section{Antenna Terms}

In this section we introduce the well known terms of directivity, gain and Q-factor for antennas see \eg~\cite{Balanis2005,Vandenbosch2010}. The aim is to precisely define the optimization problems that we study in the next section. Consider an antenna of finite extent placed in free space at the origin of our coordinate system. The antenna far-field radiation intensity $U(\hr)$ varies with the unit-direction $\hr$ as seen from the center of the antenna. The total radiated power $\rP$ of the antenna is defined as
\begin{equation}
\rP = \int_{\Ss^2} U(\hr)\diff \Omega = \frac{1}{2\eta} \lim_{R_0\rightarrow \infty }\int_{|\vr|=R_0} |\vE(\vr)|^2\diff S
\end{equation}
where $\Ss^2$ is the unit sphere in $\RR^3$, and $\diff \Omega=\sin\theta\diff \theta\diff \phi$ and $\diff S=r^2\diff \Omega$ and  where we let $\theta$ be the polar angle and $\phi$ is the azimuth angle. Here $\eta$ is the free space impedance. Given the currents densities on the antenna, we can directly determine its electric field, $\vE$, and hence its radiated power. Here, $\vr$, is a vector in $\RR^3$ and $r=|\vr|$ with $\hr=\vr/r$. It is often convenient to express the radiation intensity in terms of the far-field $\vF(\hr)$ where $r\vE(\vr)\lexp{\ju kr} \rightarrow \vF(\hr)$ as $r\rightarrow \infty$. We have that $U(\hr)=|\vF(\hr)|^2/(2\eta)$.

The total directivity, $D(\hr)$, of an antenna is defined as the ratio of the radiation intensity in the direction $\hr$ to the average radiation intensity, $\bar{U}$, over the sphere, \ie
\begin{equation}
D(\hr) = \frac{U(\hr)}{\bar{U}} = \frac{4\pi U(\hr)}{\rP},
\end{equation}
and the peak total directivity is $\max_{\hr\in \Ss^2} D(\hr)$. 

The gain, $G(\hr)$, of an antenna accounts also for the efficiency $\delta=\rP/P_{\text{in}}$, in the antenna where $P_{in}$ is the input power to the antenna. The gain is defined as
\begin{equation}
G(\hr)=D(\hr)\delta.
\end{equation}
The partial gain, and the partial directivity are related to the directivity of a given polarization of the radiated field. That is consider a polarization-direction, $\he$, orthogonal to the direction of observation direction $\hr$, then the partial directivity (partial-gain) is the radiation intensity in that polarization of the field \ie  $U(\hr,\he)=|\he\cdot\vF(\hr)|^2/(2\eta)$ to the average radiated power (input power).  In the present paper we consider only loss-less antennas, \ie $\delta=1$, where gain and directivity are equal, the reason for this limitation is that we are concerned here mainly with antennas of metal with very low losses. Indeed lossy antennas tend to have a larger bandwidth, and less radiated power due to the presence of Ohmic losses in the materials. Thus it is a tougher problem to obtain a large bandwidth in the loss-less case.

An interesting class of antennas are superdirective antennas see \eg~\cite{Hansen2006,Yaghjian+etal2008}. It is well known that infinite directivity is possible~\cite{Oseen1922,Uzkov1946,Riblet1948,Hansen1981}. Superdirectivity, is somewhat vaguely defined as antennas with above normal directivity, as defined in \eg~\cite{Harrington1958,Geyi2003}. However, since the directivity is unbounded it is essential to compare at what cost the directivity comes. The essential feature investigated here is bandwidth, since size does not necessarily limit large directivity. However, a small size does limit the $Q$-factor, making it more expensive in terms of bandwidth to increase the directivity.  

Small antenna designs tend to have a low directivity. However, a higher directivity can be obtained at the expense of antenna bandwidth. An advantage with small antennas, \eg\ antennas with $ka<1$, where $k$ is the wave number and $a$ is the radius of a sphere that circumscribe the antenna, are that the Q-factor gives an implicit way to define their relative bandwidth~\cite{Yaghjian+Best2005,Gustafsson+Nordebo2006}. The fractional (or relative) bandwidth $B$ for a single resonance circuit is related to the Q-factor through the relation~\cite{Yaghjian+Best2005} near a single resonance:
\begin{equation}
B = \frac{f_2-f_1}{f_0}\approx \frac{2\Gamma_0}{Q\sqrt{1-\Gamma_0^2}},
\end{equation}
where the center frequency $f_0=(f_1+f_2)/2$ and the magnitude of the maximally allowed reflection coefficient is $\Gamma_0$. When the input impedance of the antenna $Z_{\text{in}}=R_\text{in}+\ju X_\text{in}$ is known one often define the impedance $Q$-factor through the relation~\cite{Yaghjian+Best2005,Gustafsson+Nordebo2006}
\begin{equation}
Q_{Z} = \frac{\sqrt{(\omega R'_\text{in})^2+(\omega X'_\text{in}+|X_\text{in}|)^2}}{2R_\text{in}}.
\end{equation}
A comparison between bandwidth and different definitions of  Q-factors is given in~\cite{Gustafsson+Jonsson2014}. {Different Q-factors give rather similar results with some deviations near regions of closely spaced multiple-resonances. The region of validity of $Q_Z$ was investigated in~\cite{Gustafsson+Nordebo2006,Stuart+etal2007}, where the latter indicate that $Q_Z$ larger than $\sim$ 2 often suffices to predict the fractional bandwidth.}

The definition of the lower Q-bound for a loss-less antenna is defined as
\begin{equation}\label{Qdef}
Q = \min_{\vJ} \frac{2\omega \max(\We(\vJ),\Wm(\vJ))}{\rP(\vJ)},
\end{equation}
where $\vJ$ is the current density on the antenna. Here $\We$ and $\Wm$ are the stored electric and the stored magnetic energy, defined below.

Different definitions of stored energies for antennas have been discussed throughly in the literature, {see \eg~\cite{Collin+Rothschild1964,Fante1969,Geyi2003,Yaghjian+Best2005,Pozar2009,Vandenbosch2010,Gustafsson+etal2012a,Capek+etal2013,Gustafsson+Jonsson2014,Jonsson+Gustafsson2015,Capek+Jelinek2015b,Geyi2015,Jonsson+Gustafsson2016,Capek+etal2016}}. Here we use the definition:
\begin{equation}\label{theW}
\We=\We^{(0)}+\Wem, \ 
\Wm=\Wm^{(0)}+\Wem,
\end{equation}
where
\begin{multline}
\We^ {(0)}= \frac{\mu}{4k^ 2}\cdot \\\int_S\int_{S} (\nabla_1\cdot \vJ_{1})(\nabla_2\cdot\vJ_{2}^*)\frac{\cos(k|\vr_1-\vr_2|)}{4\pi |\vr_1-\vr_2|}\diff S_1\diff S_2,
\end{multline}
\begin{equation}
\Wm^ {(0)}= \frac{\mu}{4}\int_S\int_{S} \vJ_{1}\cdot\vJ_{2}^*\frac{\cos(k|\vr_1-\vr_2|)}{4\pi |\vr_1-\vr_2|}\diff S_1\diff S_2,
\end{equation}
and
\begin{multline}
\Wem=\frac{-\mu}{4k}\int_S\int_S \big(k^2\vJ_{1}\cdot\vJ_{2}^ *-(\nabla_1\cdot\vJ_{1})(\nabla_2\cdot \vJ_{2}^*)\big)\cdot\\\frac{\sin(k|\vr_1-\vr_2|)}{8\pi}\diff S_1 \diff S_2.
\end{multline}
{Above we introduced the notation $\vr_1,\vr_2\in \RR^3$, $\vJ_1=\vJ(\vr_1)$ and similarly for $\vJ_2$, $\mu$ is the free space permeability. The surface $S$ is the surface of the antenna structure, or a surface that enclose the structure. The $j=1,2$ in the integration element $\diff S_j$ indicate that the integration is with respect to the coordinates $\vr_j=(x_j,y_j,z_j)$, $j=1,2$.} 

This definition of stored energies~\eqref{theW} is valid for antennas without magnetic material. Stored energy for antennas with magnetic materials are investigated in~\cite{Yaghjian+etal2013,Jonsson+Gustafsson2015,Jonsson+Gustafsson2016,Kim2016}. We have furthermore made the simplification that the current densities $\vJ$ are surface current densities.  {The stored energies that have been proposed, see \eg~\cite{Yaghjian+Best2005,Vandenbosch2010,Capek+Jelinek2015b,Gustafsson+Jonsson2014,Yaghjian+etal2013} differs slightly in definition and in value. It was also observed that some of the proposed energies had a weak coordinate-dependence for a discussion see~\eg~\cite{Jonsson+Gustafsson2015,Jonsson+Gustafsson2016}, where it also was shown that such a dependence is a perturbation of higher order for small $ka$. The here used stored energies are coordinate independent.}

We note that the Q-factor in~\eqref{Qdef} depends on the set of allowed current densities. If we make the space of allowed currents larger, by \eg\ increasing the support of current densities we find a lower Q-factor. Thus a shape that enclose an antenna have a lower Q-factor bound, than the antenna.

The stored energies and the radiated power as well as the power intensity are determined numerically utilizing a Method of Moment (MoM) approach. Note that the stored energies are very similar to the components of the electric field integral equations, and it is thus easy to extract them from a MoM code. The in-house MoM-code is based on the  Rao-Wilton-Glisson (RWG) basis functions $\{\vpsi_n(\vr)\}_{n=1}^N$ and use a 10:th order Gaussian quadrature~\cite{Dunavant1985} to determine the impedance matrix. For the singular terms in the impedance matrix we use DECDEM as described in~\cite{Polimeridis+Mosig2011} and references therein. 
Thus, current densities $\vJ$ are approximated by the (current) coefficients $I=(I_1,I_2,\ldots,I_N)$ so that
\begin{equation}
\vJ \approx \sum_{n=1}^N I_n \vpsi_n(\vr).
\end{equation}
Utilizing the Galerkin-method, to obtain the finite dimensional approximation (MoM) of the electric field integral equations, we find that the electric stored energy can be expressed as $2\omega \We\approx\dotp{I}{\we I}$, where $\we$ is a finite $N\times N$-matrix acting upon the current density coefficient vector $I$.  Similarly we also have $2\omega \Wm\approx\dotp{I}{\wm I}$. 

The radiated field is estimated by $\rP\approx\dotp{I}{\rp I}$. Here we use the notation $\dotp{x}{y}=\sum_{n=1}^N x_n^*y_n$ as the inner product between two vectors with $N$-elements. The electric far-field $\vF(\hr)$ in a given polarization $\he$ and $\hr$ is approximated through 
\begin{multline} 
\he\cdot \vF(\hr) = 
\frac{-\ju k \eta}{4\pi} \int_S \he^*\cdot \vJ(\vr_1)\lexp{\ju k\hr\cdot \vr_1}\diff S \approx \\
\frac{-\ju k \eta}{4\pi} \sum_{n=1}^N I_n \int_S \he^*\cdot \vpsi_n(\vr_1)\lexp{\ju k\hr\cdot \vr_1}\diff S =
\dotp{f(\hr,\he)}{I},
\end{multline}
where $f$ is the $[N,1]$-vector with components proportional to the integral of the conjugate of the polarization vector with the RWG-basis as shown above. In the excellent tutorial~\cite{Gustafsson+etal2016a} one can find a discussion on both matrix formulations of key antenna terms as well as its application to convex optimization. 

\section{Bounds on the Q-factor}

In this section we formulate three different minimization problems towards the goal to determine a lower bound on the Q-factor for a given directivity-constraint. 

\subsection{Minimization of the Q-factor}

Lets start with the bounds on the Q-factor for a given partial directivity. In~\cite{Gustafsson+Nordebo2013} they formulated a $D/Q$-bound for superdirectivity, utilizing the observation that 
\begin{equation}
\Dp\leq D(\hr,\he)\approx \frac{4\pi|\dotp{f(\hr,\he)}{I}|^2}{2\eta\dotp{I}{\rp I}}.
\end{equation}
The maximum of the $D/Q$ problem at $D\geq \Dp$ can be reformulated as a convex minimization problem~\cite{Gustafsson+Nordebo2013}:
\begin{equation}\tag{P}\label{pQ}\begin{aligned}
\minimize_{I\in \CC^N} & \max(\dotp{I}{\we I},\dotp{I}{\wm I})\\
\subjectto & \dotp{f(\hr,\he)}{I} = -\ju,\\
& \dotp{I}{\rp I} \leq \frac{2\pi}{\eta \Dp}.
\end{aligned}\end{equation}
We refer to \eqref{pQ} as the problem of finding the Q-factor for a given {\it partial} directivity $D(\hr,\he)$. 
Here $\CC^N$ is $N$-vectors with complex coefficients. Maximizing $D(\hr,\he)/Q$ with constraints on $D(\hr,\he)$, naturally gives $Q$ at a given $D(\hr,\he)\geq \Dp$. 

Observe that the minimization reformulation of the problem does not state that we have obtained the peak partial directivity in the given direction $\hr$, only that the directivity in the desired direction is larger than $\Dp$. The ingenious idea in~\cite{Gustafsson+Nordebo2013} to formulate Q-factor bounds with a given partial directivity is easily solvable with cvx~\cite{Grant+Boyd2014} or another convex-optimization solver given the respective matrices $\we,\wm$ and $\rp$ together with the far-field vector $f(\hr,\he)$. The results yields both a minimizing current on the structure, as well as the unique minimum of the problem, see \eg~\cite{Gustafsson+etal2016a}.

If we instead want to formulate bounds on the Q-factor for the total directivity in the direction $\hr$, \eg\ $\min Q$ for $D(\hr)\geq \Dr$, we need to include both polarizations of the far-field \eg\ $\he$, $\hh$. Here $(\hr,\he,\hh)$ forms a right-hand triple at each observation direction $\hr$. We note that $|\dotp{f(\hr,\he)}{I}|^2=\dotp{I}{f(\hr,\he)f^H(\hr,\he)I}$. Lets introduce the matrix 
\begin{equation}\label{um}
   \um(\hr)=\frac{2\pi}{\eta}(f(\hr,\he)f^H(\hr,\he)+f(\hr,\hh)f^H(\hr,\hh)),
\end{equation}
which is related to the radiation intensity through $U(\hr)\approx \dotp{I}{\um(\hr) I}/(4\pi)$. The approximation is as usual due to that we have only a finite number of RWG-basis functions in the representation of $U(\hr)$. 

Given the matrix $\um$ in~\eqref{um}, we find that the total directivity can hence be estimated through 
\begin{equation}
\Dr\leq D(\hr)\approx \frac{\dotp{I}{\um(\hr) I}}{\dotp{I}{\rp I}}.
\end{equation}
Note that the optimization problem $\min Q$ given the total directivity $D(\hr)\geq \Dr$ is scaling invariant in the amplitude of the current density. The numerical approximation of this minimization problem can hence be formulated as the following non-convex problem 
\begin{equation}\tag{Q}\label{tQ}\begin{aligned}
\minimize_{I\in \CC^N} & \max(\dotp{I}{\we I},\dotp{I}{\wm I})\\
\subjectto & \dotp{I}{\rp I} = 1,\\
& \dotp{I}{\um(\hr) I} \geq \Dr.
\end{aligned}\end{equation}
We refer to the problem~\eqref{tQ} as the problem of finding the Q-factor for a given {\it total} directivity $\Dr$. Clearly, given a solution current $I_*$, we find the $Q_*$ as a function of $D_*(\hr)\geq \Dr$. 

It is interesting to compare the solution to the lower Q-bounds in the respective case of \eqref{pQ} and \eqref{tQ}, since it is not always easy to predict which polarization that gives the lowest Q-factor. It depends both on the shape of the object, but also on the desired observation direction, see \eg\ Fig~\ref{fig1} in next section. The information that there are possible currents that provide a lower Q-factor if we relax our demand on the polarization is important, since the possibility of such currents that radiate part or all of its power in another polarization-direction will affect the antenna design. If polarization purity is important, then design measures to suppress the unwanted polarization may be required. However in \eg\ harvesting applications it is an advantage if multiple polarizations are absorbed by the antenna. Thus the cross-polarization levels are often an important factor in antenna designs. 

The minimization problem given in~\eqref{tQ} is non-convex, and has to the authors knowledge not been re-formulated as a convex optimization problem in the literature. Observe that this optimization problem fall in the category of quadratically constrained quadratic optimization programs (QCQP). Beyond linear programming, quadratic programming is a rather well investigated class of optimization problems. QCQP-type problems are often occurring in physics applications where \eg\ energies are quadratic forms in terms of sources. There are several different methods to solve~\eqref{tQ}, either directly with a non-linear or heuristic solver or through a relaxation method. A powerful method to relax QCQP into convex problems is the semi-definite relaxation method, SDR. To derive the relaxed problem we follow the SDR-technique and observe that 
\begin{equation}
\dotp{I}{\rp I} = \tr(I^H\rp I)=\tr(\rp \xm),
\end{equation}
where $\xm=II^H$ is a rank 1 matrix and $\tr(\xm)$ is the trace of the matrix $\xm$. With this observation we can now formulate the optimization problem for the lower Q-bound given a total directivity in the $\hr$-direction as
\begin{equation}\tag{R}\label{rQ}
\begin{aligned}
\minimize_{0\leq \xm \in \HH_+^N} & \max(\tr(\we \xm),\tr(\wm \xm))\\
\subjectto & \tr(\rp \xm) = 1,\\
& \tr(\um(\hr) \xm) \geq \Dr,
\end{aligned}
\end{equation}
where we have dropped, relaxed, the $\rank \xm=1$ condition in order to make the problem convex. Here $\HH_+^N\subset\CC^{N\times N}$ is the set of $N\times N$ hermitian positive semi-definite matrices. The problem \eqref{rQ} falls in the class of semi-definite programming, and there are efficient methods to solve rather large such problems see \eg~\cite{ODonoghue+etal2016}. The problem stated in~\eqref{rQ} is the relaxed problem in finding the lowest Q-factor for a {\it total} directivity, in a given direction. 

If we compare these different minimization problems for bounds the Q-factor we note that both \eqref{pQ} and \eqref{rQ} are convex and hence each has a unique minimum. However~\eqref{rQ} has a much larger set of unknowns in its $N\times N$-matrix $\xm$ that is Hermitian positive semi-definite, as compared with, the $N$-vector of $I$. We thus tend to obtain rather large size minimization problems for~\eqref{rQ}. The relaxed problem \eqref{rQ} will always be a lower bound to the original problem \eqref{tQ}. In fact, in our case we have a stronger statement, which is that \eqref{rQ} is tight and predicts the solution to~\eqref{tQ} in all cases. This fact is shown in next subsection. 

We also observe that a numerically calculated solution $\xm_*$ might not be a rank one matrix. There are however several methods to extract feasible rank one solutions, see \eg~\cite{Luo+etal2010}.

Above we have focused on the total directivity. We note in passing that several other desirable constraints \eg\ like antenna tuning $\We=\Wm$, {embedding of antennas~\cite{Gustafsson+etal2016a} as the ground plane of a cellular phones}, losses, and other constraints also can be formulated as relaxed convex problems since they all fall in the category of quadratically constrained quadratic programming. See also~\cite{Shi+etal2017,Jonsson+etal2017,Ehrenborg+Gustafsson2017} for additional applications.

\subsection{The trace-minimum is the solution to the original problem}\label{Sp}

A feature of the SDR technique is that the relaxed problem gives a rather tight approximation of the originally investigated problem see \eg~\cite{Luo+etal2010}. Indeed under some circumstances it determines the minimum to the original solution. 

To show that \eqref{rQ} is tight, we observe that $\min Q$ for $D(\hr)\geq \Dr$ is equivalent to $\max D(\hr)$ for $Q\leq \Qz$, when the latter problem is feasible. If we express this latter maximization problem in terms of the matrices, once again using the scaling invariance, we find that it can be written as: 
\begin{equation}\label{mD}
   \begin{aligned}
      \maximize_{I\in \CC^N} & \dotp{I}{\um(\hr) I} \\
      \subjectto & \dotp{I}{\rp I} = 1,\\
         & \dotp{I}{\we I} \leq \Qz, \\
         & \dotp{I}{\wm I} \leq \Qz.
   \end{aligned}
\end{equation}
We have thus established that solving~\eqref{mD} is equivalent to solving \eqref{tQ} if the solution is feasible. Note that there is a minimal $Q$-factor, $\Qm$ associated with a bounded structure. For small antennas it was shown in~\cite{Jonsson+Gustafsson2015} that $\Qm=4\pi/(k^3 \gamma)$, where $\gamma$ is the maximal eigenvalue of the electric polarizability matrix, in our case. Thus, for $\Qz\geq \Qm$ there is a solution to the above problem. The trace-relaxed version of \eqref{mD} is
\begin{equation}\label{Dr}
   \begin{aligned}
      \maximize_{0\leq \xm\in \HH_+^{N}} & \tr (\um(\hr) \xm) \\
      \subjectto & \tr(\rp \xm)) = 1,\\
         & \tr(\we \xm)\leq \Qz, \\
         & \tr(\wm \xm)\leq \Qz.
   \end{aligned}
\end{equation}
Both \eqref{mD} and \eqref{Dr} have three constraints\footnote{An approach to~\eqref{rQ} is to replace $\minimize \max(\dotp{I}{\we I},\dotp{I}{\wm I})$ with $\minimize t$, with constraints $\dotp{I}{\we I}\leq t$, and $\dotp{I}{\wm I}\leq t$. Thus~\eqref{tQ} has four constraints, resulting in that~\cite{Huang+Zhang2007} predict $\rank \xm_*\leq 2$ for its solutions. It is therefor beneficial to rewrite it into a problem for $\max D$ as given in~\eqref{mD} and \eqref{Dr}.} and from a theorem in~\cite[\S5]{Huang+Zhang2007} we deduce that there are minimizers $\xm_*$ such that $\rank \xm_*=1$. Note that the result in~\cite{Huang+Zhang2007} does not say that all solutions to~\eqref{Dr} have rank one, only that there are such solutions if the problem is feasible. Following the argumentation in the proof of~\cite[Thm.~5]{Luo+etal2007} we note that since $\rank \xm_*=1$ there is an $I_*$ such that $\xm_*=I_* I_*^H$, and this is the optimal current density solution to~\eqref{mD}, since \eqref{Dr} is always a upper bound to~\eqref{mD}. 

Since the problems~\eqref{mD} and \eqref{tQ} are equivalent for $\Qz\geq \Qm$, we have that both problems are tight. That is, the minimum of \eqref{rQ} is also the minimum of~\eqref{tQ}.

\section{A Generalized Eigenvalue-Approach to Q-factor Bounds}

In this section we introduce an alternative method to determine the lower bounds on the  Q-factor for a total directivity~\eqref{tQ}. We use this latter method to illustrate that the solution to the relaxed problem~\eqref{rQ} indeed predict the solution to the original problem~\eqref{tQ}. We also use this method to solve problems where the memory demands of solution methods to~\eqref{rQ} become too large. The method derived here have some similarities with the method given in~\cite{Jelinek+Capek2017}.

It was observed in~\cite{Gustafsson+etal2016a} that for $\alpha\in [0,1]$
\begin{equation}\label{a}
\max(\dotp{I}{\we I},\dotp{I}{\wm I})\geq \dotp{I}{(\alpha \we + (1-\alpha)\wm)I},
\end{equation}
which was used in~\cite{Capek+etal2016b} to show that the duality gap for Q-factors are tight, \ie that $\wa = \alpha \we + (1-\alpha)\wm$ can be used in a generalized eigenvalue problem to determine the Q-factor bound without any constraints. That is the problem
\begin{equation}\label{uQ}
\begin{aligned}
\minimize_I & \dotp{I}{\wa I}\\
\subjectto & \dotp{I}{\rp I}=1,
\end{aligned}
\end{equation}
can be formulated as a generalized eigenvalue problem $\wa I = \lambda_{\alpha} \rp I$, see also~\cite{Jonsson+Gustafsson2015} for the operator version. Lower bounds on $Q$ are shown in~\cite{Capek+etal2016b} to be equal to $Q=\max_\alpha \min_{n\in N} \lambda_{\alpha,n}$.  
Certain care has to be applied numerically since the matrix $\rp$ can fail to be numerically positive definite even though $\rP>0$. This is associated with that $\rp$ is a low rank matrix corresponding to the propagating vector spherical modes of the antenna, for a discussion see~\cite{Gustafsson+etal2016a}. 

To generalize~\eqref{uQ} to include constrains on the total directivity is straightforward, we find
\begin{equation}\label{aQ}\begin{aligned}
\minimize_{I\in \CC^N} & \dotp{I}{\wa I}\\
\subjectto & \dotp{I}{\rp I} = 1,\\
& \dotp{I}{\um(\hr) I} \geq \Dr.
\end{aligned}\end{equation}
The Lagrangian with Lagrange multipliers $(\lambda,\sigma)$ associated with~\eqref{aQ} is 
\begin{equation}\label{LL}
\LL = \dotp{I}{[\wa - \lambda (\rp-1) + \sigma (\um(\hr)- \Dr)]I}.
\end{equation}
The Karush-Kuhn-Tucker conditions associated with~\eqref{aQ} and its dual are given as condition on the solutions $(\sigma,\lambda,I)$ to~\eqref{aQ}:
\begin{align}~\label{EL}
(\wa +\sigma \um(\hr))I = \lambda \rp I, \\
   \sigma (\dotp{I}{\um(\hr) I}-\Dr)=0, \ \ \sigma\leq 0\\
   \dotp{I}{\um(\hr) I}\geq \Dr, \ \ \dotp{I}{\rp I}=1.
\end{align}
One method to find solutions to this problem could consist of the following steps:
\begin{itemize}
   \item Fix the parameters $\alpha$ and $\sigma$ and determine all generalized eigenvalues $\{\lambda_n\}$ and associated eigenvectors $I_n$ to~\eqref{EL}. 
   \item Sweep $\alpha\in [0,1]$ and $\sigma\in (-\infty,0]$ to find the optimal solution $(I,\lambda)$ to~\eqref{aQ} among all investigated parameters $(\alpha,\sigma)$.  
\end{itemize}
It is a slow search towards the appropriate optimal solution. The slowness is due to both the large $N$ and the $N\times N$ generalized eigenvalue problem in~\eqref{EL} which has to be solved a large number of times to trace out the feasible space in the parameters $(\alpha,\sigma)$. 

Certain care has to be take, given the additional parameter $\sigma$ as compared to the problem given in~\eqref{uQ}, since $\sigma$ corresponds to the directivity constraint. It is not clear as of yet if this approach can be simplified similarly to~\cite{Capek+etal2016} to give the unique minimizer. There is the possibility that~\eqref{aQ} has a duality gap as compared to~\eqref{tQ}.

One approach to circumvent such a duality gap is found by the observation that any current $I$ inserted in the original problem \eqref{tQ} is an upper approximation of the lowest bound. The search approach followed here to find the lowest possible bound is as follows: 
\begin{itemize}
\item Utilize solutions $(\lambda_n,I_n)$ to the generalized eigenvalue problem given in~\eqref{EL} for a particular $(\alpha,\sigma)$. 
\item Select the sub-set of feasible solutions, \ie these that satisfies the minimal directivity constraints. Test which of these ones give the smallest $Q$-factor in~\eqref{tQ}.
\item Repeat the above procedure for a new sample of $(\alpha,\sigma)$. 
\item After an exhaustive search over some grid of the $(\alpha,\sigma)$-space, we use the best solution as an initial point in matlab's fminsearch over $(\alpha,\sigma)$ to further approach the minimum. 
\end{itemize}
The method is rather slow, however we can speed it up by noticing that both $\um$ and $\rp$ are of low rank, which substantially reduces the set of eigenvalues that needs to be investigated at each $(\alpha,\sigma)$.

\section{Numerical Examples -- Q-factor Bounds for Superdirective Antennas}

\subsection{Single plate structure} 
\begin{figure}[thbp]
  \centering
  \includegraphics[width=\linewidth]{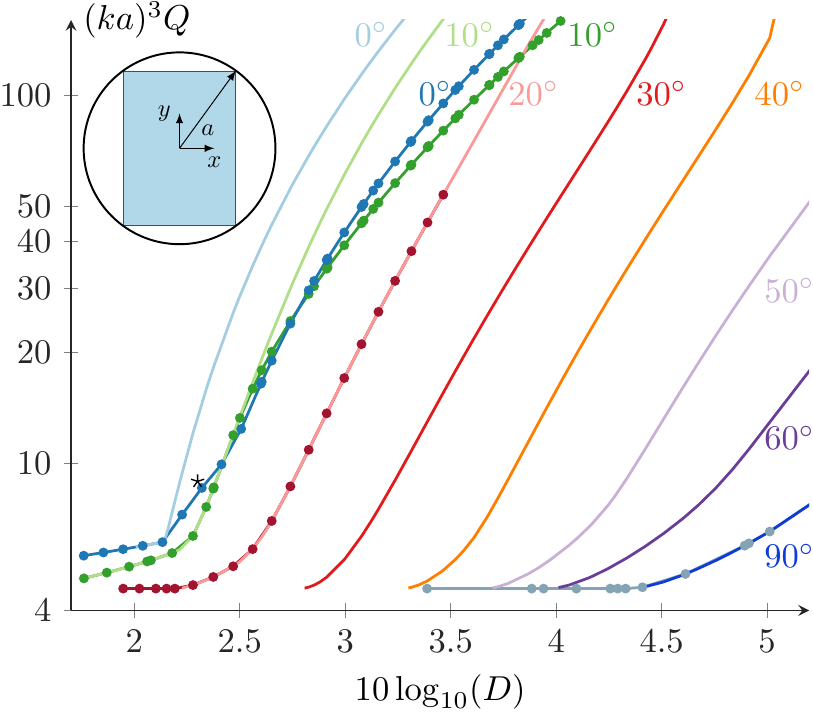}
   \caption{Comparison between optimization~\eqref{pQ} with partial directivity and trace-optimization~\eqref{rQ}. The trace-optimization curves are lines marked with small dots and are depicted above for $\theta\in[0^\circ,10^\circ,20^\circ,90^\circ]$. For $\theta\in[20^\circ-90^\circ]$ both minimizations yield the same minimum in the considered range of Q-values, apart from that the trace-minimization find solutions also for lower directivity, as is indicated as the horizontal-tail of $\theta=90^\circ$ in light blue, in the lower right-hand corner of the graph. The star at $D=2.3$ dBi is a full wave simulation $(D,Q_Z)$ at 0.855 GHz of one of the elements depicted in Fig.~\ref{fig3}. {The plate has $ka=0.51$ at 0.9 GHz.} The $(ka)^3$ scaling of $Q$ allows us to compare results for frequencies where the antenna is small. See \eg\ the comparisons of \eqref{pQ} for different frequencies in Fig.~\ref{fig2}.
}
  \label{fig1}
\end{figure}

Here we solve the optimization problems~\eqref{pQ},~\eqref{tQ} and \eqref{rQ} while sweeping the $\hr(\theta,\phi)$-direction for the polar angle $\theta\in [0^\circ,90^\circ]$ and $\phi=0^\circ$. The shape is a rectangle depicted in the insert in Fig.~\ref{fig1}. That is we determine an optimal current on an infinitesimally thin rectangle of size 32 mm $\times$ 44 mm centered in the xy-plane, at the frequency $f=0.9$ GHz, giving the electrical size $ka=0.51$. We compare the result between the problem \eqref{pQ} for $\he=\hphi=\hy$ and the trace-Q minimization~\eqref{rQ} that allows both polarizations. We find that the trace-minimization has a lower Q-bound for $\theta\leq 10^\circ$ as compared with the partial directivity optimization problem~\eqref{pQ}, in particular for $D>2.1$ and $D>2.5$ (dBi) respective for $\theta=0^\circ$ and $\theta=10^\circ$.
At the investigated directions with $\theta\geq 20^\circ$, we note that the trace-optimization~\eqref{rQ} and~\eqref{pQ} optimization largely agree for the range of $Q$-values considered. The rectangle has a lowest possible $Q$-value $\Qm$, determined without directivity constraints. We note that the trace-minimization recaptures $\Qm$ for low enough directivity, as illustrated with the flat start on the $\theta=90^\circ$ curve with trace-minimization. 

The non-linear solver agrees with the trace-minimization problem. This illustrates the in~\S\ref{Sp} established fact, that in the investigated case we do not only obtain a lower estimate of the Q-value, but the solution to the optimization problem~\eqref{tQ}. The envelope that consists of $(ka)^3Q\approx 4.6$ for $D<3.5$ dBi, combined with the $\theta=90^\circ$-curve is an estimate of the lower range of the bound of the Q-factor as a function of the total  directivity in this range of $(D,Q)$. 

{The total directivity over all directions $\hr$ for an antenna is defined as $D_\mrm{m}=\max_{\hr} D(\hr)$, clearly $D_{\mrm{m}}\geq D(\hr)$. Above we use~\eqref{tQ} to determine $Q$ for a desired $\Dr$ and direction $\hr$. Fig.~\ref{fig2} is determined by subsequent optimization over different $\hr$'s. The distinction between $D_\mrm{m}$ and $D(\hr)$, also explains the appearance of low-directivity solutions in Fig.~\ref{fig1}. That is, $D(\hr)\leq 2$ dBi imply that in a certain direction $\hr$ it is possible to obtain a very low directivity. This is a natural phenomena, think about an elementary dipole antenna with the direction chosen along the dipole. Explicit examples are given \S\ref{sec:twoplate}, where max directivity $D_\mrm{m}$ is compared with the realized directivity for a current-density optimized for in a given direction $\hr$.}

As always in this kind of simulations it is possible to increase the resolution and thus decrease the minimal $Q$-value to a small extent. In the present calculation we have used a $N= 10\times 14$ regular grid on the rectangle as input to the Delaunay triangularization of the surface, resulting in an impedance matrices with $\sim 400 \times 400$ elements.

\begin{figure}[htbp]
  \centering
  \includegraphics[width=0.7\linewidth]{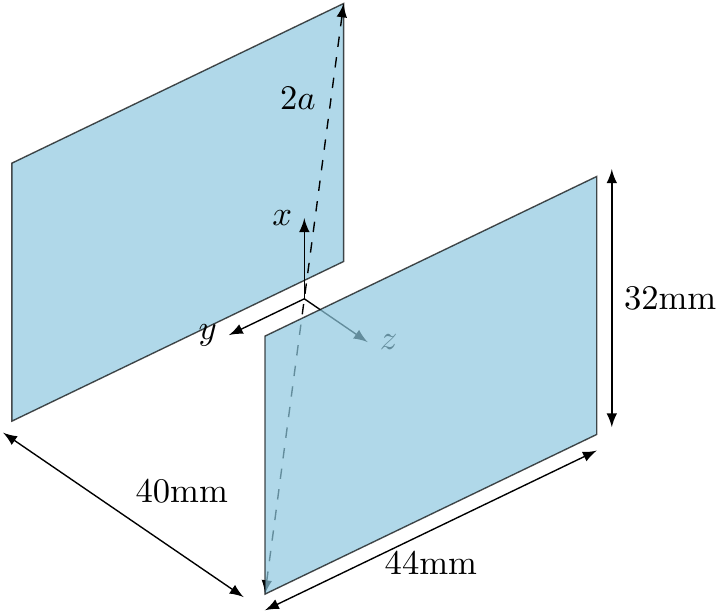}
  \caption{The structure used in the second example, see Fig.~\ref{rad} and Fig.~\ref{fig2}. 
}
  \label{IIshape}
\end{figure}

\subsection{Two plate structure} \label{sec:twoplate}
\begin{figure}
  \centering
  \includegraphics[width=\linewidth]{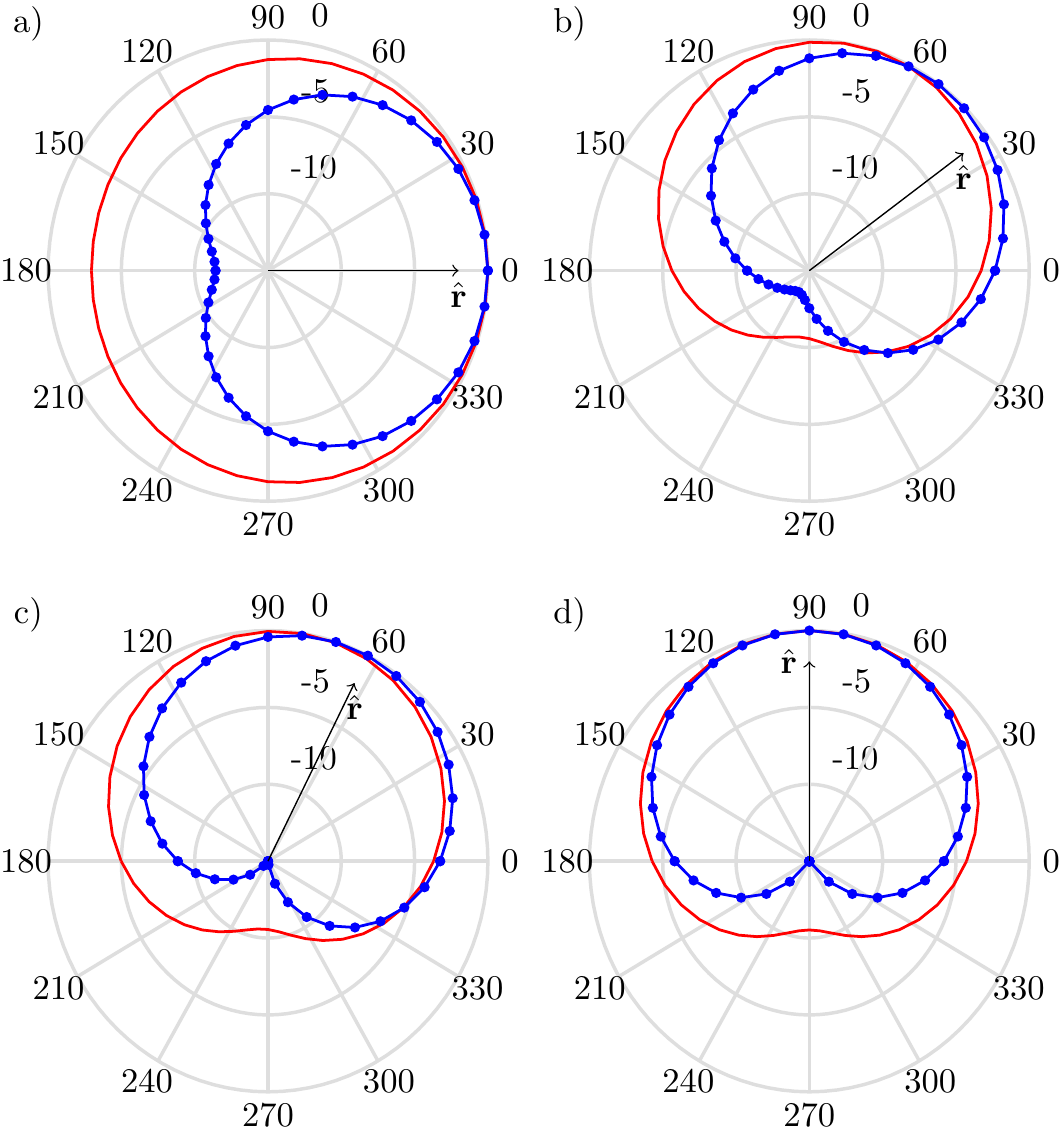}
  \caption{The normalized radiation patterns in [a,b,c,d] above correspond to a $\hr(\theta,\phi)$-goal direction indicated in the figures with an arrow pointing in the directions $\theta\in [0^\circ,36^\circ,63^\circ,90^\circ]$, $\phi=0$. The red line above corresponds to the lowest $D$-value possible solving~\eqref{pQ}, smallest Q-factor of the blue lines in Fig~ \ref{fig3} with $\theta=[0^ \circ,36^ \circ,63^\circ,90^ \circ]$. The blue lines, with dots are radiation patterns with $D(\hr)=\Dr\approx 5.4$ dBi for the same desired $\hr$-directions.
}
  \label{rad}
\end{figure}

In our second example, we consider two rectangular plates of 32 mm $\times$ 44 mm, parallel to the xy-plane and with a distance between them of 40 mm along the z-axis, see Fig.~\ref{IIshape}. The optimization problem is solved for a range of different directions $\hr(\theta,\phi)$. The polar angle $\theta$ is chosen in the range $\theta\in [0^\circ,90^\circ]$, and the azimuthal angle is $\phi=0$. We use $f=0.9$ GHz if not otherwise stated. The directional constraints in the optimization is not equivalent with that the peak directivity is in that direction, as mentioned above. To illustrate this we utilize the convex minimization in~\eqref{pQ} at the lowest possible $D$ for the $\hr(\theta,\phi)$ angles $\theta\in[0^\circ,36^\circ,63^\circ,90^\circ]$. The radiation pattern in the $xz$-plane with a $y$-polarized field correspond to the lowest Q-value for each of these angles are depicted in red in Fig.~\ref{rad}. The radiation patterns marked with a blue line with dots corresponds to the same desired $\theta$-angles in $\hr$, but now at directivity $D\approx 5.4$ dBi. The realized radiation pattern at the two intermediate angles do not have their peak directivity at the aimed for angles of $36^\circ$ and $63^\circ$ respectively. We note that the optimization does not require that the peak directivity is in the direction of $\hr$. However, as the requested directivity $D(\hr)$ becomes larger, we see that the main peak-direction approaches the desired direction $\hr$. See Fig.~\ref{rad}b, and~\ref{rad}c, where the peak of the blue line moves towards the goal $\hr$-direction. 

\begin{figure}
  \centering
  \includegraphics[width=\linewidth]{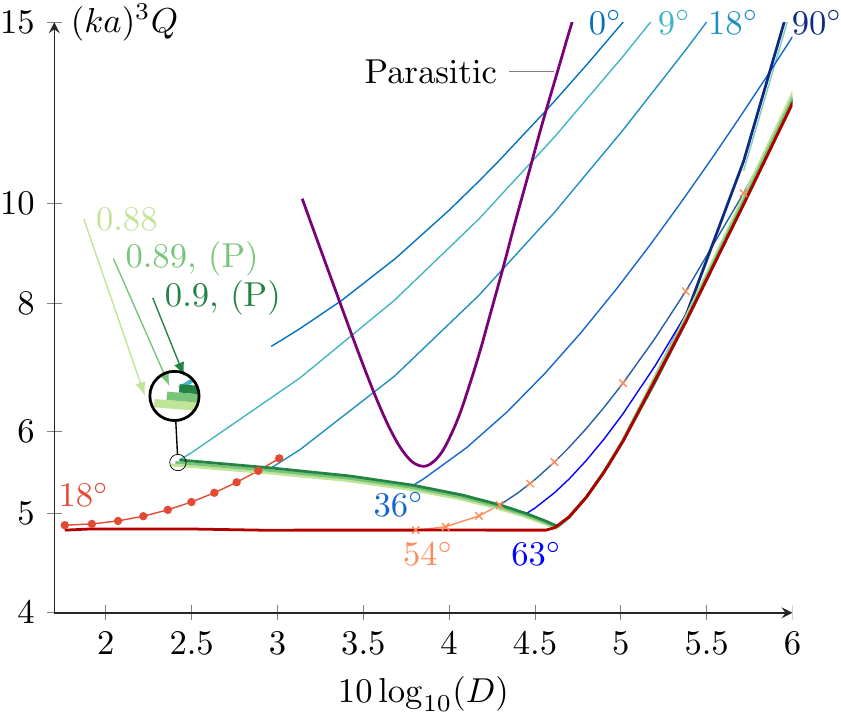}
   \caption{Comparison between optimization problem \eqref{pQ} with partial directivity, trace-optimization~\eqref{rQ} and eigenvalue search~\eqref{tQ}. The trace-optimization curves at $f=0.9$ GHz are marked with small dots and are depicted above for $\theta=18^\circ$ and $N=8\times 11$ (orange). The partial directivity cases \eqref{pQ} are sampled in blueish colors for $\theta\in[0^\circ,9^\circ,18^\circ,36^ \circ,54^\circ,63^\circ,90^\circ]$, and $N=10\times 14$. The eigenvalue search is included for $\theta=54^\circ$. It also defines the dark-red envelope at the bottom of the figure corresponding to the lowest $Q$ sweeping $\theta\in[0^\circ,90^\circ]$. The green-hue colors corresponds to the envelope of the lowest $Q$-values from the partial directivity optimization \eqref{pQ} for frequencies [0.88, 0.89, 0.9] GHz. {The structure has the electrical size $ka=0.64$ @ 0.9 GHz.} The line marked parasitic corresponds to a Parasitic element antenna, see Fig.~\ref{fig3} and~\cite{Ferrero+etal2017}. 
}
  \label{fig2}
\end{figure}

In Fig.~\ref{rad}b, at the lowest $Q$-value (the red curve) we find that the peak radiation direction is $\approx 70^\circ$, instead of the expected $36^\circ$. Our interpretation of this solution to the minimization problem~\eqref{pQ} is that the current minimizer consists mainly of a current with low-Q radiation at $90^\circ$ perturbed by a current that ensures that the direction $36^\circ$ has large enough far-field amplitude. These currents are scaled relative to each other in such a way so that the constraints in~\eqref{pQ} are satisfied. The interpretation is hence that it is more costly in Q-factor (smaller bandwidth) to have the peak directivity at $36^\circ$ than it is to scale up a slightly perturbed radiation pattern with peak directivity at $90^\circ$ sufficiently much to satisfy the constraint $\dotp{f(\hr,\he)}{I}=-\ju$ resulting in a peak radiation direction at $\approx 70^\circ$. By increasing the demands on $\Dr$ to $D\approx 5.4$ dBi we find that the $Q$-factor also increases, see Fig.~\ref{fig2}. This ensures that it costs more in Q-factor to use a peak directivity in another direction, since high directivity amounts to less radiation in non-peak directions. This is also indicated with the blue line in Fig.~\ref{rad}b where the peak direction moves more towards $36^\circ$. We see the same phenomena for a desired peak direction at $63^\circ$ in Fig.~\ref{rad}c. 

{The above illustrated case also helps to explain the solutions with directivity below 2.3 dBi. Solutions to the~\eqref{tQ}-problem in this directivity range have their peak directivity direction in a different direction than the desired $\hr$-direction, utilizing both available polarizations. } 
 
\begin{figure}
  \centering
  \includegraphics[width=\linewidth]{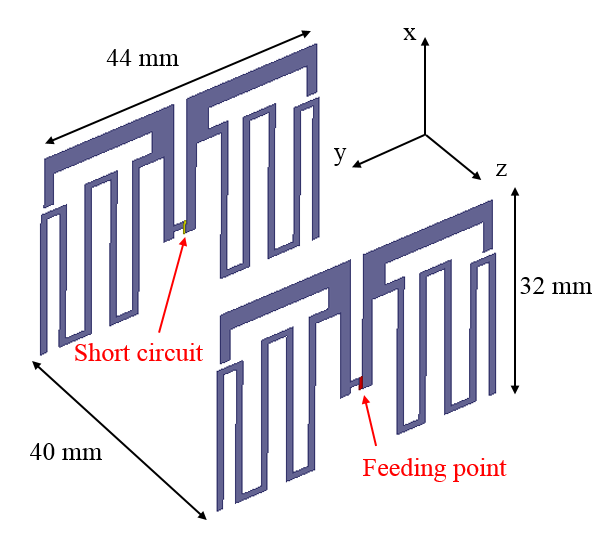}
  \caption{EM Model of the two plates parasitic superdirective antenna.
}
  \label{fig3}
\end{figure}

The lower bounds on Q for a given partial directivity~\eqref{pQ} are depicted in blueish colors in Fig.~\ref{fig2}. The smallest in Q-factor envelope over all~\eqref{pQ} optimizations upon sweeping over the desired observation direction $\theta$ is shown in green. The green lines consists of the frequencies $f=[0.88,0.89,0.90]$ GHz. Due to the $(ka)^3$-normalization of the $Q$-factor in~\eqref{fig3} we find that they essentially are coinciding, showing that the electric polarization is essential in determining these $Q$-bounds~\cite{Jonsson+Gustafsson2015}. 

The red envelope in Fig.~\ref{fig2} is the smallest $Q$ in solving \eqref{tQ} using the generalized eigenvalue method. For Q-values that is larger than the green line, we have that the lowest $Q$-value for~\eqref{pQ} and \eqref{tQ} agree for the considered range, as illustrated with $\theta=54^\circ$-case. The mesh used in this case is the same as the mesh for one rectangle, \ie $N=10\times 14$ for each rectangle. 

To investigate the stability of the solutions with respect to increasing the mesh size we also have used $N=15\times 21$, with only a few points that are inserted below and almost coinciding with the line for $\theta=90^\circ$ up in the upper right corner of Fig.~\ref{fig2} (light blue). 

Comparing \eqref{pQ} and \eqref{tQ} with solutions to the trace-optimization~\eqref{rQ} is harder in this case. Solutions with $N=5\times 7$ and $N=8\times11$ fill the region between the red and the green line, with small perturbations, as illustrated with the $\theta=18^\circ$ for $N=8\times 11$. To illustrate a bit of the challenge in solving the $N=10\times 14$-case with cvx, we find that the SeDuMi solver converts and solves the problem of size dimension $\sim 4.8\cdot 10^5$ for the $N=8 \times 11$-case, whereas the $N=10\times 14$ correspond to a problem of size $1.2\cdot 10^6$. This rapid growth limits the size of the investigated problems. However we note that a comparison between the speed of the trace-optimization and the non-linear eigenvalue search shows that the trace-method is faster. Thus given sufficient memory, trace-optimization is an attractive method. We once again observe that the trace-bounds on $Q$ is tight, \ie they are also solutions to \eqref{tQ}. 

Another phenomena that is illustrated in Fig.~\ref{fig2} is that the lowest possible $Q$-factor for a particular directivity is not necessarily one with radiation in a cardinal direction. Note that at $D>5.3$ dBi we see that the curve marked $90^\circ$ has a higher Q-value than the curves marked $\theta=54^\circ$ and $63^\circ$. In~Fig.~\ref{rad}c we have that the blue line have its peak directivity at $\approx 63^\circ$. Similarly the blue radiation pattern with $D\approx 5.4$ dBi in Fig.~\ref{rad}d has its peak at $90^\circ$. Thus, an interpretation is that as we solve the minimization problem  with higher requirement on the directivity, we find current minimizers also with non-cardinal peak radiation direction with a lower Q-factor than currents that radiate in the cardinal direction. It is nice to see how this process still smoothly trace out the lower-Q envelope as a function of $D$ given by the low-Q red line in Fig.~\ref{fig2}. In essence, we find that depicted curve furthermost to the right (red) is a description of the lowest Q-factor increase with demands on directivity. The red line thus describe the smallest cost in Q-factor for a desired (super)directivity. 
 
In Fig.~\ref{fig2} we have also included a lilac curve marked `Parasitic' that corresponds to a full-wave simulation over a frequency interval close to the resonance of a high-directivity parasitic element antenna of the shape shown in Fig~\ref{fig3}. The shape fits into the investigated double-rectangle structure. It is composed with a driven element (lower plate) and a parasitic element (upper plate). It is closely related to a realized and measured antenna, discussed in~\cite{Ferrero+Lizzi2016,Ferrero+etal2017}. We see that the lowest point of the parasitic curve at the frequency 0.892 GHz comes rather close to the low green~\eqref{pQ}-case \eg\ it has a fairly high directivity, while keeping the Q-factor low. If it is possible to realize antennas on the green-curve, or even better closer the red marked lower Q-factor envelope in~\eqref{fig2} is still an open question. 

\section{Conclusion}

We show that semi-definite relaxation can be used to determine the Q-factor for a given total directivity. We have shown that it often agrees with the Q-bounds for a given {\it partial} directivity. However, in both the illustrated cases we also note that there are situations with lower Q-factors than predicted by the earlier method if we allow for both polarizations. 

The reason that we use the lowest Q-factor to bound (super)directivity is that size does not necessarily limit the directivity. However, as we have illustrated, given a desired Q-factor, there is a limited maximal directivity.  Thus by connecting directivity to its maximally allowed Q-factor for a given shape, we find an approximation to the widest bandwidth associated with a desired total directivity. 

Another result shown in the paper is that the semi-definite relaxation method give tight bounds on the original quadratically constrained problem of determining the lowest Q-factor at a given total directivity $D(\hr)$. Thus SDR formulation of the matrix version of $\max D(\hr)$ for $Q\leq \Qz$, $Q_0 \geq \Qm$ is tight.

\section*{Acknowledgements}

LJ and FF are grateful to the Labex UCN@Sophia for having funded Lars Jonsson visiting period in University Nice Sophia.
LJ gratefully acknowledge the support form the Swedish foundation for strategic research for the project ``Convex analysis and convex optimization for EM design", and LJ and LW gratefully acknowledge the support from Swedish Governmental Agency for Innovation Systems through the center ChaseOn in the project iAA. SS gratefully acknowledge funding from CSC.


\begin{thebibliography}{10}
\providecommand{\url}[1]{#1}
\csname url@samestyle\endcsname
\providecommand{\newblock}{\relax}
\providecommand{\bibinfo}[2]{#2}
\providecommand{\BIBentrySTDinterwordspacing}{\spaceskip=0pt\relax}
\providecommand{\BIBentryALTinterwordstretchfactor}{4}
\providecommand{\BIBentryALTinterwordspacing}{\spaceskip=\fontdimen2\font plus
\BIBentryALTinterwordstretchfactor\fontdimen3\font minus
  \fontdimen4\font\relax}
\providecommand{\BIBforeignlanguage}[2]{{%
\expandafter\ifx\csname l@#1\endcsname\relax
\typeout{** WARNING: IEEEtran.bst: No hyphenation pattern has been}%
\typeout{** loaded for the language `#1'. Using the pattern for}%
\typeout{** the default language instead.}%
\else
\language=\csname l@#1\endcsname
\fi
#2}}
\providecommand{\BIBdecl}{\relax}
\BIBdecl

\bibitem{Le+etal2016}
T.~N. Le, A.~Pegatoquet, T.~Le~Huy, L.~Lizzi, and F.~Ferrero, ``Improving
  energy efficiency of mobile wsn using reconfigurable directional antennas,''
  \emph{IEEE Communications Letters}, vol.~20, no.~6, pp. 1243--1246, 2016.

\bibitem{Gustafsson+Nordebo2013}
M.~Gustafsson and S.~Nordebo, ``Optimal antenna currents for {Q},
  superdirectivity, and radiation patterns using convex optimization,''
  \emph{IEEE Trans. Antennas Propagat.}, vol.~61, no.~3, pp. 1109--1118, 2013.

\bibitem{Gustafsson+etal2016a}
M.~Gustafsson, D.~Tayli, C.~Ehrenborg, M.~Cismasu, and S.~Nordebo, ``Antenna
  current optimization using {MATLAB} and {CVX},'' \emph{{FERMAT}}, vol.~15,
  no.~5, pp. 1--29, 2016.

\bibitem{Capek+etal2016b}
\BIBentryALTinterwordspacing
M.~Capek, M.~Gustafsson, and K.~Schab, ``Minimization of antenna quality
  factor,'' \emph{ArXiv e-prints, 1612.0767}, 2016. [Online]. Available:
  \url{http://arxiv.org/abs/1612.0767}
\BIBentrySTDinterwordspacing

\bibitem{Yaghjian+Best2005}
A.~D. Yaghjian and S.~R. Best, ``Impedance, bandwidth, and {$Q$} of antennas,''
  \emph{IEEE Trans. Antennas Propagat.}, vol.~53, no.~4, pp. 1298--1324, 2005.

\bibitem{Gustafsson+Jonsson2014}
M.~Gustafsson and B.~L.~G. Jonsson, ``Antenna {Q} and stored energy expressed
  in the fields, currents, and input impedance,'' \emph{IEEE Trans. Antennas
  Propagat.}, vol.~63, no.~1, pp. 240--249, 2015.

\bibitem{Oseen1922}
C.~W. Oseen, ``Die {E}insteinsche {N}adelstichstrahlung und die {M}axwellschen
  {G}leichungen,'' \emph{Annalen der Physik}, vol. 374, no.~19, pp. 202--204,
  1922.

\bibitem{Uzkov1946}
A.~Uzkov, ``An approach to the problem of optimum directive antenna design,''
  in \emph{Comptes Rendus (Doklady) de l’Academie des Sciences de l’URSS},
  vol.~53, no.~1, 1946, pp. 35--38.

\bibitem{Riblet1948}
H.~Riblet, ``Note on the maximum directivity of an antenna,'' \emph{Proceedings
  of the IRE}, vol.~36, no.~5, pp. 620--623, May 1948.

\bibitem{Hansen1981}
R.~C. Hansen, ``Fundamental limitations in antennas,'' \emph{Proc. IEEE},
  vol.~69, no.~2, pp. 170--182, 1981.

\bibitem{Harrington1958}
R.~Harrington, ``On the gain and beamwidth of directional antennas,''
  \emph{IEEE Trans. Antennas Propagat.}, vol.~6, no.~3, pp. 219--225, 1958.

\bibitem{Geyi2003}
W.~Geyi, ``Physical limitations of antenna,'' \emph{IEEE Trans. Antennas
  Propagat.}, vol.~51, no.~8, pp. 2116--2123, Aug. 2003.

\bibitem{Lovasz1979}
L.~Lov{\'a}sz, ``On the shannon capacity of a graph,'' \emph{IEEE Transactions
  on Information theory}, vol.~25, no.~1, pp. 1--7, 1979.

\bibitem{Goemans+Williamson1995}
M.~X. Goemans and D.~P. Williamson, ``Improved approximation algorithms for
  maximum cut and satisfiability problems using semidefinite programming,''
  \emph{Journal of the ACM (JACM)}, vol.~42, no.~6, pp. 1115--1145, 1995.

\bibitem{Hansen2006}
R.~C. Hansen, \emph{Electrically Small, Superdirective, and Superconductive
  Antennas}.\hskip 1em plus 0.5em minus 0.4em\relax New Jersey: John Wiley \&
  Sons, 2006.

\bibitem{Yaghjian+etal2008}
A.~D. Yaghjian, T.~H. O'Donnell, E.~E. Altshuler, and S.~R. Best,
  ``Electrically small supergain end-fire arrays,'' \emph{Radio Science},
  vol.~43, no.~3, pp. 1--13, 2008.

\bibitem{Lo+Lee1966}
Y.~Lo, S.~Lee, and Q.~Lee, ``Optimization of directivity and signal-to-noise
  ratio of an arbitrary antenna array,'' \emph{Proceedings of the IEEE},
  vol.~54, no.~8, pp. 1033--1045, 1966.

\bibitem{Margetis+etal1998}
D.~Margetis, G.~Fikioris, J.~M. Myers, and T.~T. Wu, ``Highly directive current
  distributions: General theory,'' \emph{Physical Review E}, vol.~58, no.~2, p.
  2531, 1998.

\bibitem{Pigeon+etal2014}
M.~Pigeon, C.~Delaveaud, L.~Rudant, and K.~Belmkaddem, ``Miniature directive
  antennas,'' \emph{International Journal of Microwave and Wireless
  Technologies}, vol.~6, no.~1, pp. 45--50, 2014.

\bibitem{Grant+Boyd2014}
M.~Grant and S.~Boyd, ``{CVX}: Matlab software for disciplined convex
  programming, version 2.1,'' \url{cvxr.com/cvx}, Apr. 2014.

\bibitem{Luo+etal2010}
Z.-Q. Luo, W.-K. Ma, A.~M.-C. So, Y.~Ye, and S.~Zhang, ``Semidefinite
  relaxation of quadratic optimization problems,'' \emph{IEEE Signal Processing
  Magazine}, vol.~27, no.~3, pp. 20--34, 2010.

\bibitem{Anstreicher2009}
K.~M. Anstreicher, ``Semidefinite programming versus the
  reformulation-linearization technique for nonconvex quadratically constrained
  quadratic programming,'' \emph{Journal of Global Optimization}, vol.~43, no.
  2-3, pp. 471--484, 2009.

\bibitem{Sun+Dai2016}
C.~Sun and R.~Dai, ``An iterative method for nonconvex quadratically
  constrained quadratic programs,'' \emph{ArXiv e-print 1609.02609}, 2016.

\bibitem{Bengtsson+Ottersten2001}
M.~Bengtsson and B.~Ottersten, ``Optimal and suboptimal transmit beamforming,''
  in \emph{Handbook of Antennas in Wireless Communications}.\hskip 1em plus
  0.5em minus 0.4em\relax CRC Press, 2001, pp. 18--1.

\bibitem{Lavaei+Low2012}
J.~Lavaei and S.~H. Low, ``Zero duality gap in optimal power flow problem,''
  \emph{IEEE Transactions on Power Systems}, vol.~27, no.~1, pp. 92--107, 2012.

\bibitem{Candes+etal2015}
E.~J. Candes, Y.~C. Eldar, T.~Strohmer, and V.~Voroninski, ``Phase retrieval
  via matrix completion,'' \emph{SIAM review}, vol.~57, no.~2, pp. 225--251,
  2015.

\bibitem{Vandenberghe+Boyd1996}
L.~Vandenberghe and S.~Boyd, ``Semidefinite programming,'' \emph{SIAM review},
  vol.~38, no.~1, pp. 49--95, 1996.

\bibitem{Kokiopoulou2011}
E.~Kokiopoulou, J.~Chen, and Y.~Saad, ``Trace optimization and eigenproblems in
  dimension reduction methods,'' \emph{Numerical Linear Algebra with
  Applications}, vol.~18, no.~3, pp. 565--602, 2011.

\bibitem{Huang+Zhang2007}
Y.~Huang and S.~Zhang, ``Complex matrix decomposition and quadratic
  programming,'' \emph{Mathematics of Operations Research}, vol.~32, no.~3, pp.
  758--768, 2007.

\bibitem{Balanis2005}
C.~A. Balanis, \emph{Antenna Theory}, 3rd~ed.\hskip 1em plus 0.5em minus
  0.4em\relax New Jersey: John Wiley \& Sons, 2005.

\bibitem{Vandenbosch2010}
G.~A.~E. Vandenbosch, ``Reactive energies, impedance, and {Q} factor of
  radiating structures,'' \emph{IEEE Trans. Antennas Propagat.}, vol.~58,
  no.~4, pp. 1112--1127, 2010.

\bibitem{Gustafsson+Nordebo2006}
M.~Gustafsson and S.~Nordebo, ``Bandwidth, {$Q$}-factor, and resonance models
  of antennas,'' \emph{Progress in Electromagnetics Research}, vol.~62, pp.
  1--20, 2006.

\bibitem{Stuart+etal2007}
H.~Stuart, S.~Best, and A.~Yaghjian, ``Limitations in relating quality factor
  to bandwidth in a double resonance small antenna,'' \emph{Antennas and
  Wireless Propagation Letters}, vol.~6, 2007.

\bibitem{Collin+Rothschild1964}
R.~E. Collin and S.~Rothschild, ``Evaluation of antenna {Q},'' \emph{IEEE
  Trans. Antennas Propagat.}, vol.~12, pp. 23--27, Jan. 1964.

\bibitem{Fante1969}
R.~L. Fante, ``Quality factor of general antennas,'' \emph{IEEE Trans. Antennas
  Propagat.}, vol.~17, no.~2, pp. 151--155, Mar. 1969.

\bibitem{Pozar2009}
D.~M. Pozar, ``{New results for minimum Q, maximum gain, and polarization
  properties of electrically small arbitrary antennas},'' in \emph{Antennas and
  Propagation, 2009. EuCAP 2009. 3rd European Conference on}, March 2009, pp.
  1993--1996.

\bibitem{Gustafsson+etal2012a}
M.~Gustafsson, M.~Cismasu, and B.~L.~G. Jonsson, ``{Physical bounds and optimal
  currents on antennas},'' \emph{IEEE Trans. Antennas Propagat.}, vol.~60,
  no.~6, pp. 2672--2681, 2012.

\bibitem{Capek+etal2013}
M.~Capek, Jelinek, L., Hazdra, P., and J.~Eichler, ``The measurable {Q} factor
  and observable energies of radiating structures,'' \emph{IEEE Trans. Antennas
  Propagat.}, vol.~62, no.~1, pp. 311--318, 2014.

\bibitem{Jonsson+Gustafsson2015}
B.~L.~G. Jonsson and M.~Gustafsson, ``Stored energies in electric and magnetic
  current densities for small antennas,'' \emph{Proc. R. Soc. A}, vol. 471, no.
  2176, p. 20140897, 2015.

\bibitem{Capek+Jelinek2015b}
M.~Capek and L.~Jelinek, ``{Various Interpretations of the Stored and the
  Radiated Energy Density},'' \emph{ArXiv e-prints: 1503.06752}, Mar. 2015.

\bibitem{Geyi2015}
W.~Geyi, ``Stored energies and radiation {Q},'' \emph{IEEE Trans. Antennas
  Propagat.}, vol.~63, no.~2, pp. 636--645, 2015.

\bibitem{Jonsson+Gustafsson2016}
B.~L.~G. Jonsson and M.~Gustafsson, ``Stored energies for electric and magnetic
  current densities,'' \emph{ArXiv e-print: 1604.08572}, pp. 1--25, 2016.

\bibitem{Capek+etal2016}
M.~Capek, L.~Jelinek, and G.~A.~E. Vandenbosch, ``Stored electromagnetic energy
  and quality factor of radiating structures,'' \emph{Proc. R. Soc. A}, vol.
  472, no. 2188, p. 201604, 2016.

\bibitem{Yaghjian+etal2013}
A.~D. Yaghjian, M.~Gustafsson, and B.~L.~G. Jonsson, ``Minimum {Q} for lossy
  and lossless electrically small dipole antennas,'' \emph{Progress In
  Electromagnetics Research}, vol. 143, pp. 641--673, 2013.

\bibitem{Kim2016}
O.~S. Kim, ``Lower bounds on {Q} for finite size antennas of arbitrary shape,''
  \emph{IEEE Trans. Antennas Propagat.}, vol.~64, no.~1, pp. 146--154, 2016.

\bibitem{Dunavant1985}
D.~A. Dunavant, ``High degree efficient symmetrical gaussian quadrature rules
  for the triangle,'' \emph{International journal for numerical methods in
  engineering}, vol.~21, no.~6, pp. 1129--1148, 1985.

\bibitem{Polimeridis+Mosig2011}
A.~G. Polimeridis and J.~R. Mosig, ``On the direct evaluation of surface
  integral equation impedance matrix elements involving point singularities,''
  \emph{IEEE Antennas and Wireless Propagation Letters}, vol.~10, pp. 599--602,
  2011.

\bibitem{ODonoghue+etal2016}
\BIBentryALTinterwordspacing
B.~O'Donoghue, E.~Chu, N.~Parikh, and S.~Boyd, ``Conic optimization via
  operator splitting and homogeneous self-dual embedding,'' \emph{Journal of
  Optimization Theory and Applications}, vol. 169, no.~3, pp. 1042--1068, June
  2016. [Online]. Available: \url{http://stanford.edu/~boyd/papers/scs.html}
\BIBentrySTDinterwordspacing

\bibitem{Shi+etal2017}
S.~Shi, L.~Wang, and B.~L.~G. Jonsson, ``On {Q}-factor bounds for a given
  front-to-back ratio,'' Accepted to APS, San Diego, US, 2017.

\bibitem{Jonsson+etal2017}
B.~L.~G. Jonsson, S.~Shi, and L.~Wang, ``Relaxation of some non-convex
  constraints for {Q}-factor optimization,'' Accepted to URSI GASS, Montreal,
  Canada, 2017.

\bibitem{Ehrenborg+Gustafsson2017}
\BIBentryALTinterwordspacing
C.~Ehrenborg and M.~Gustafsson, ``Fundamental bounds on {MIMO} antennas,''
  \emph{ArXiv e-prints, 1704.06600v2}, pp. 1--5, 2017. [Online]. Available:
  \url{http://arxiv.org/abs/1704.06600v2}
\BIBentrySTDinterwordspacing

\bibitem{Luo+etal2007}
Z.-Q. Luo, N.~D. Sidiropoulos, P.~Tseng, and S.~Zhang, ``Approximation bounds
  for quadratic optimization with homogeneous quadratic constraints,''
  \emph{SIAM Journal on optimization}, vol.~18, no.~1, pp. 1--28, 2007.

\bibitem{Jelinek+Capek2017}
L.~Jelinek and M.~Capek, ``Optimal currents on arbitrarily shaped surfaces,''
  \emph{IEEE Trans. Antennas Propag.}, vol.~65, no.~1, p. 329–341, 2017.

\bibitem{Ferrero+etal2017}
\BIBentryALTinterwordspacing
F.~Ferrero, L.~Lizzi, B.~L.~G. Jonsson, and L.~Wang, ``A two-element parasitic
  antenna that approach the minimum {Q}-factor at a given directivity,''
  \emph{ArXiv e-prints, 1705.02281}, pp. 1--11, 2017. [Online]. Available:
  \url{http://arxiv.org/abs/1705.02281}
\BIBentrySTDinterwordspacing

\bibitem{Ferrero+Lizzi2016}
F.~Ferrero and L.~Lizzi, ``Super-directive parasitic element antenna for
  spatial filtering applications,'' in \emph{Antennas and Propagation
  (APSURSI), 2016 IEEE International Symposium on}.\hskip 1em plus 0.5em minus
  0.4em\relax IEEE, 2016, pp. 1753--1754.

\end{thebibliography}


\end{document}